\documentclass[12pt]{iopart}

\usepackage{graphicx}

\begin{document}

\title{Emergence of Clusters in Growing Networks with Aging}

\author{Nuno Crokidakis and Marcio Argollo de Menezes}

\address{
  Instituto de F\'{\i}sica, Universidade Federal Fluminense \\
  Av. Litor\^anea s/n \\
  24210-340, Niter\'oi - RJ, Brazil}

\ead{nuno@if.uff.br, marcio@if.uff.br}

\begin{abstract}
We study numerically a model of nonequilibrium networks where nodes
and links are added at each time step with aging of nodes and
connectivity- and age-dependent attachment of links. By varying the
effects of age in the attachment probability we find, with numerical
simulations and scaling arguments, that a giant cluster emerges at a
first-order critical point and that the problem is in the universality
class of one dimensional percolation.  This transition is followed by
a change in the giant cluster's topology from tree-like to
quasi-linear, as inferred from measurements of the average
shortest-path length, which scales logarithmically with system size in
one phase and linearly in the other.

\end{abstract}

\noindent

\maketitle


\section{INTRODUCTION}

The understanding of natural, technological and social phenomena
through the network perspective has motivated a large body of research
in the past $10$ years \cite{mendesrmp,barabasi_review,mendesbook}, as
many of their properties can be inferred from the simple idea that
relations among elements in such systems can be interpreted, at a
certain level of abstraction, as nodes and links of a complex network.
Examples range from physically interacting proteins in the cell
\cite{hawoong,prot_review} and the set of routers comprising the
Internet \cite{faloutsos} to social networks \cite{eu,watts} and,
according to empirical measurements of a few network properties, like
degree distribution and correlations, clustering and average shortest
path, it is observed that many of these networks are strikingly
similar (in the statistical sense) \cite{barabasi_review}.  Thus,
generic features of these networks can be predicted by the analysis of
simple models.  Since the classic work of Erd\"os and R\'enyi
\cite{er1,er2}, a whole set of interesting results in percolation
theory have been brought up revealing the importance of network
structure in determining critical properties
\cite{callaway,mendes_pre64,molloy1,molloy2,strogatz,cohen} . In
particular, it has been found that in nonequilibrium (growing)
networks with exponential \cite{callaway} or power-law degree
distribution \cite{mendes_pre64} an infinite order critical point
separates a phase with many finite clusters from another where a
single macroscopic connected cluster emerges. It has been shown both
numerically \cite{callaway} and analytically \cite{mendes_pre64} that
all the derivatives of the average size of the largest cluster, taken
as the order parameter, are zero at the critical point. Here we study
a similar problem, where networks are grown by the addition of nodes
and links that attach preferentially, but with an age-dependent
probability. With computer simulations and scaling arguments we show
that one has a first-order transition in the size of the giant
connected cluster as one makes it less likely to attach links to older
nodes. Following the transition the topology of the largest cluster
changes from tree-like to one-dimensional, as one can infer from
measurements of the average shortest-path length inside the largest
cluster in each phase.


\section{THE MODEL}

Let a network grow from an initial cluster of $m_0$ fully-connected
nodes (we use $m_0=2$) by the addition of a node and a link at each
time step.  This new link will randomly join a pair of nodes $(i,j)$
with probability
$\Pi(k_i,k_j,a_i,a_j,t)=\Pi(k_i,a_i,t)\Pi(k_j,a_j,t)$, where
$\Pi(k,a,t)=C(A_0+k)e^{-\alpha (t-a)}$ is the probability that a node
with $k$ links and added at time $a\le t$ to the network receives a
link \footnote{an initial attractiveness \cite{mendes_prl} $A_0>0$ is
  necessary for the newly added node, which has $k=0$, to participate
  in the dynamics. We set $A_0=1$ in the following.} and
$C(t)=\sum_{i}\Pi(k_i,a_i,t)$.  Self-links and multiple links between
the same pair of nodes are forbidden. The parameter $\alpha$ plays the
role of the inverse of a timescale $\beta$, which suppresses the
attractiveness of older nodes for new connections. The case $\alpha=0$
($\beta\to \infty$) has been extensively studied in
\cite{mendes_pre64}, who reported an infinite-order phase transition
as the number $b$ of links added per time step is varied around a
critical value $b_c$. Here we analyze the effect of varying the memory
parameter $\alpha$ and find another nontrivial phase transition at a
critical value $\alpha_c$ separating a phase with multiple clusters
from another where a single, ${\cal O}(N)$, giant component emerges.

\begin{figure}[!htbp]
\vspace{0.3cm}
\begin{center}
\includegraphics[width=0.4\columnwidth,angle=0]{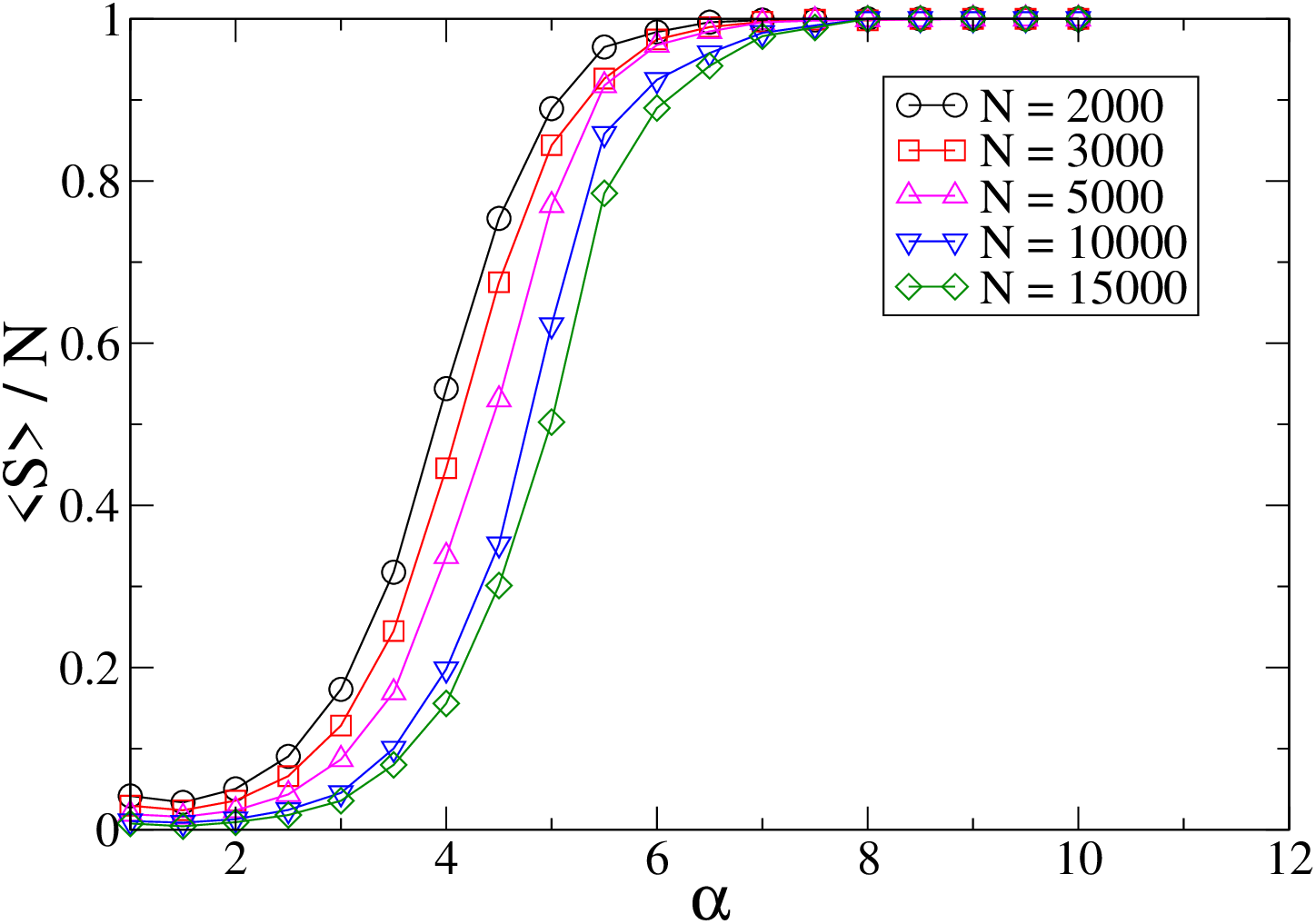}
\hspace{0.3cm}
\includegraphics[width=0.4\columnwidth,angle=0]{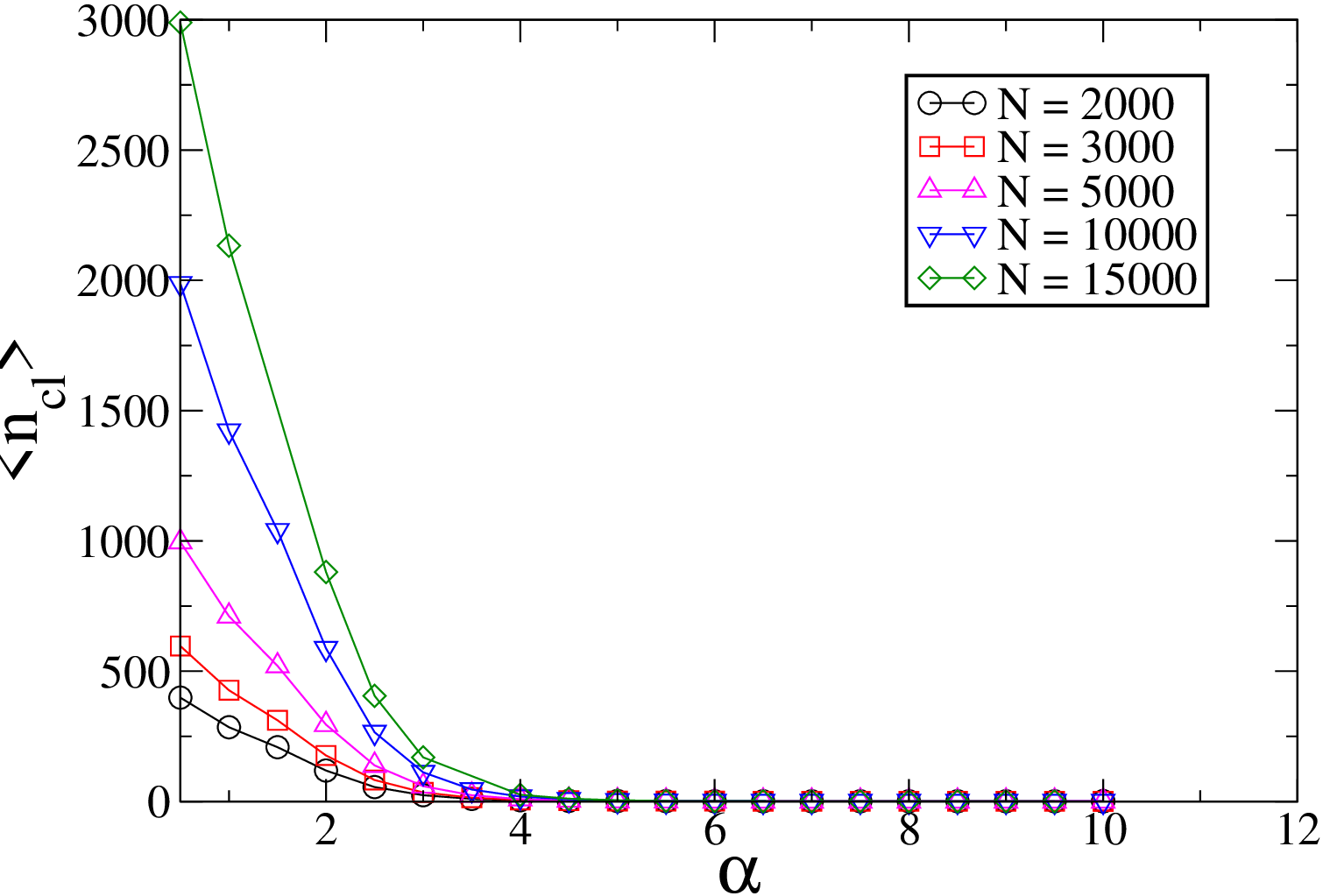}
\end{center}
\caption{The relative size of the largest cluster (left) and the
  average number of clusters (right) as a function of $\alpha$. We
  observe a phase transition at a given value $\alpha_c$ where one
  giant cluster emerges.  Results are averaged over $10^{3}$ samples
  and clusters are identified with breadth-first-search.}
\label{Fig1}
\end{figure}

For each value of $\alpha$ we grow $10^3$ networks with up to
$N=15000$ nodes. Since we do not force the newly added node to
immediately attach to a preexisting cluster, there is always a
possibility to create networks with isolated clusters.  For a given
value of $\alpha$, we identify these clusters in each network with
breadth-first-search and calculate the average size of the largest
cluster $\langle S(\alpha)\rangle$ in each network, normalized by the total
number of nodes $N$ (Fig. \ref{Fig1}).

\begin{figure}[!ht]
\begin{center}
\vspace{0.5cm}
\includegraphics[width=0.4\columnwidth,angle=0]{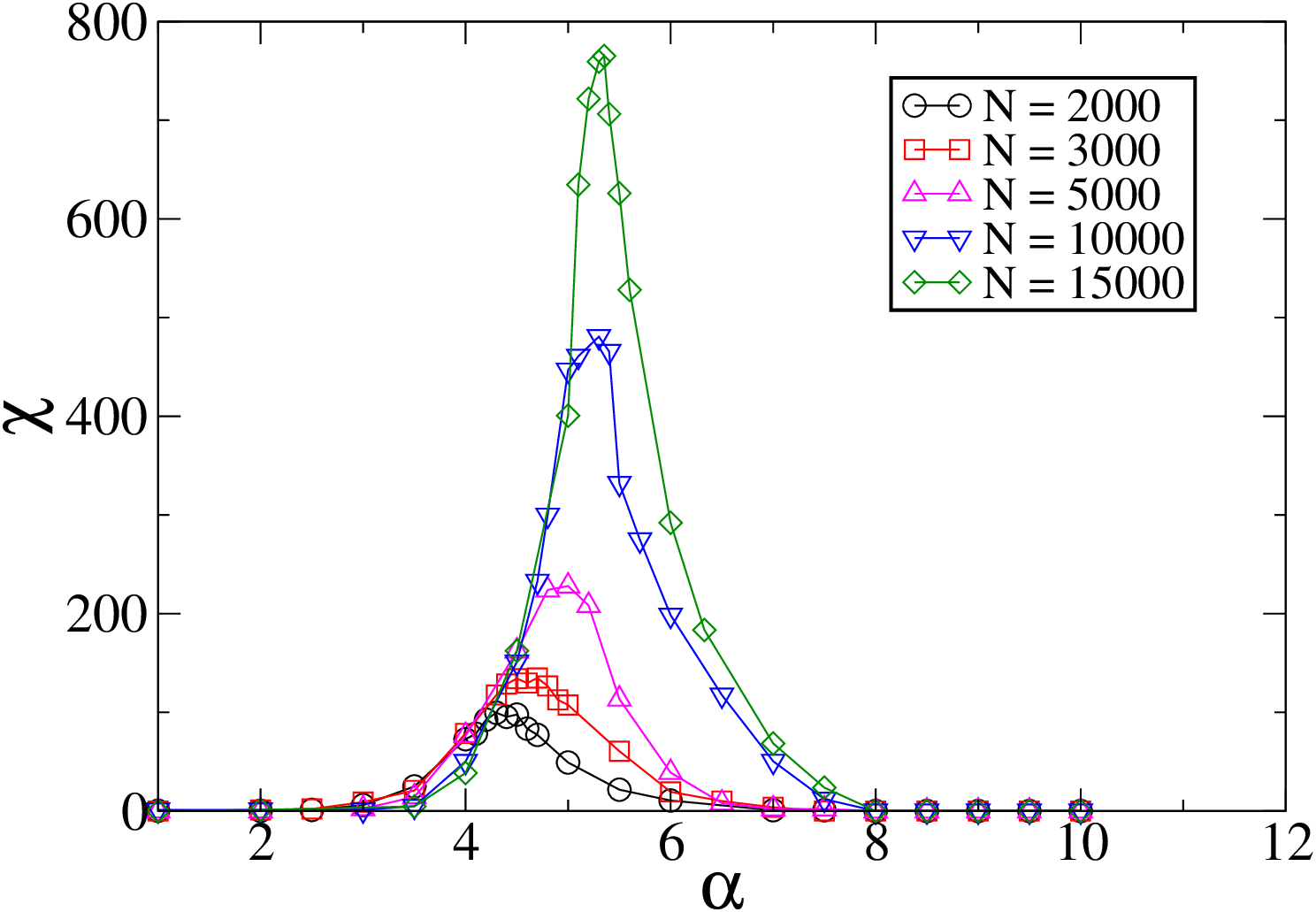}
\hspace{0.5cm}
\includegraphics[width=0.4\columnwidth,angle=0]{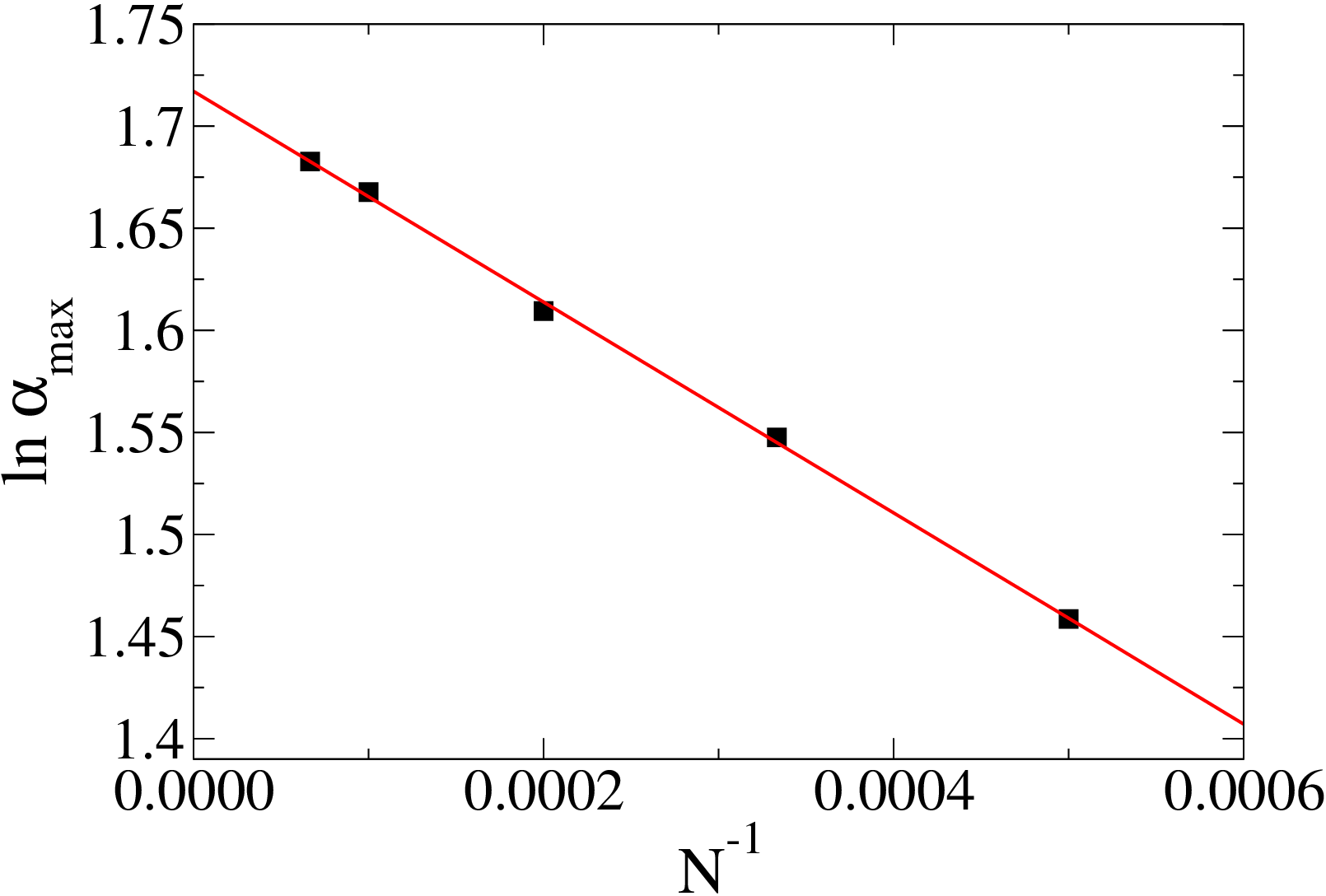}
\end{center}
\caption{Fluctuation of the order parameter for different network
  sizes as a function of $\alpha$ (left) and the position of their
  maxima $\alpha_{c}(N)$ versus the inverse of  network size $N^{-1}$
  (right). The straight line corresponds to the best fit
  $\ln\;y=1.7171-516.17\;x$.}
\label{Fig2}
\end{figure}

For small values of $\alpha$ the network is fragmented in many small
clusters, whereas after a (size-dependent) value $\alpha_c(N)$ a giant
component most likely exists. In order to determine the value of
$\alpha_c(N)$ we analyze the fluctuations of the order parameter
$\chi(\alpha,N)=\langle S^2\rangle-\langle S\rangle^2$ as a function of $\alpha$ and identify the position of the maximum of $\chi$ with $\alpha_c(N)$
(Fig. \ref{Fig2} on the left). Finite-size scaling analysis of usual
(first or second order) phase transitions suggests a power-law scaling for the
critical point shift
\begin{equation}
\label{eq1}
\alpha_c(N)=\alpha_c(\infty)+A~ N^{-1/\nu},
\end{equation}

\noindent where $A$ is a constant and $\nu=1/d$ at a first-order
(discontinuous) phase-transition \cite{fisher}. Nevertheless, we find
that for this model the position of the maxima of the
susceptibility-like parameter $\chi(\alpha_c(N),N)$ scale as
\begin{equation}
\label{eq2}
\ln \alpha_c(N)=\ln \alpha_c(\infty) - K\,N^{-1},
\end{equation}

\noindent
where $\ln \alpha_c(\infty)=1.717(1)$ and $K=516.1(7)$.  In the
thermodynamic limit ($N\to\infty$) we obtain $\alpha_{c}= 5.568(2)$ (We refer to $\alpha_c(\infty)$ as $\alpha_c$ in the following).

\begin{figure}[!ht]
\begin{center}
\vspace{1.0cm}
\includegraphics[width=0.5\columnwidth,angle=0]{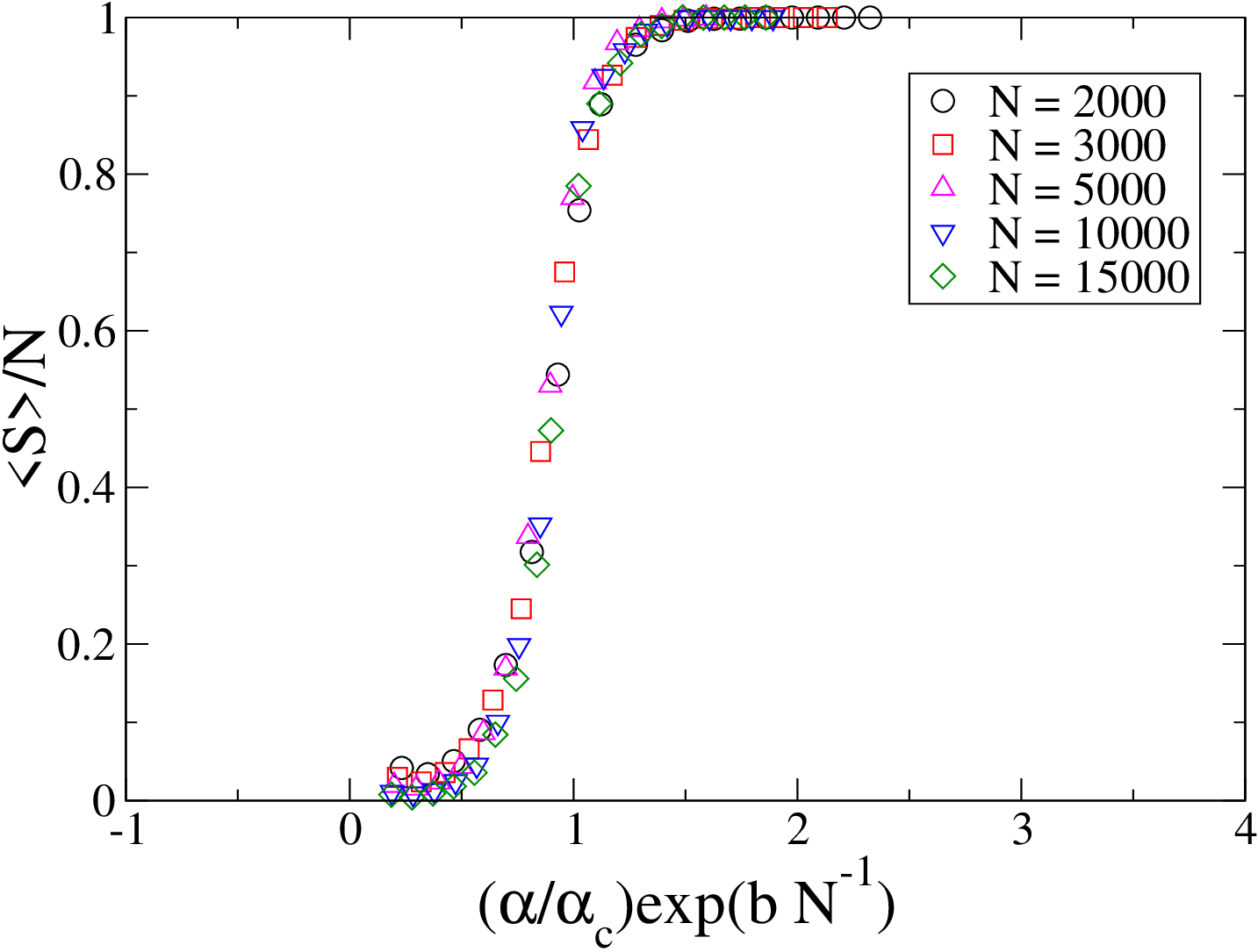}
\end{center}
\caption{Scaling plot of the order parameter $<S>/N$ as a function of
  $\frac{\alpha}{\alpha_c(N)}$, where
  $\alpha_c(N)=\alpha_c e^{-bN^{-1}}$. The values of the parameters
  are $\alpha_{c}=5.5682$ and $b=516.17$.}
\label{Fig3}
\end{figure}

As another unusual feature of this phase transition, we find that the
relative size of the largest cluster $\langle S\rangle/N$ is a function of the
ratio $\alpha/\alpha_c(N)$, as can be seen in Fig. \ref{Fig3}. This
might result from the fact that, close to $\alpha_c$, the
dimensionless scaling variable for the average size of the largest
cluster $\langle S\rangle$ is $x=\log(\alpha)$, as appears from the
scaling of the rounding of the phase transition [Eq. (\ref{eq2})], so a
function of distance in variables $x$ translates into a function of
the ratio in variables $\alpha$. It is noteworthy to mention that the
same occurs in $1d$ percolation \cite{stauffer}, where the
characteristic length $\xi(p)$ scales as $\log(p/p_c)^{-1}$ which can be
expanded as $\xi(p) \approx |p-p_c|^{-1}$ when $p$ is close enough to
$p_c=1$.

\begin{figure}[!ht]
\begin{center}
\vspace{1.0cm}
\includegraphics[width=0.5\columnwidth,angle=0]{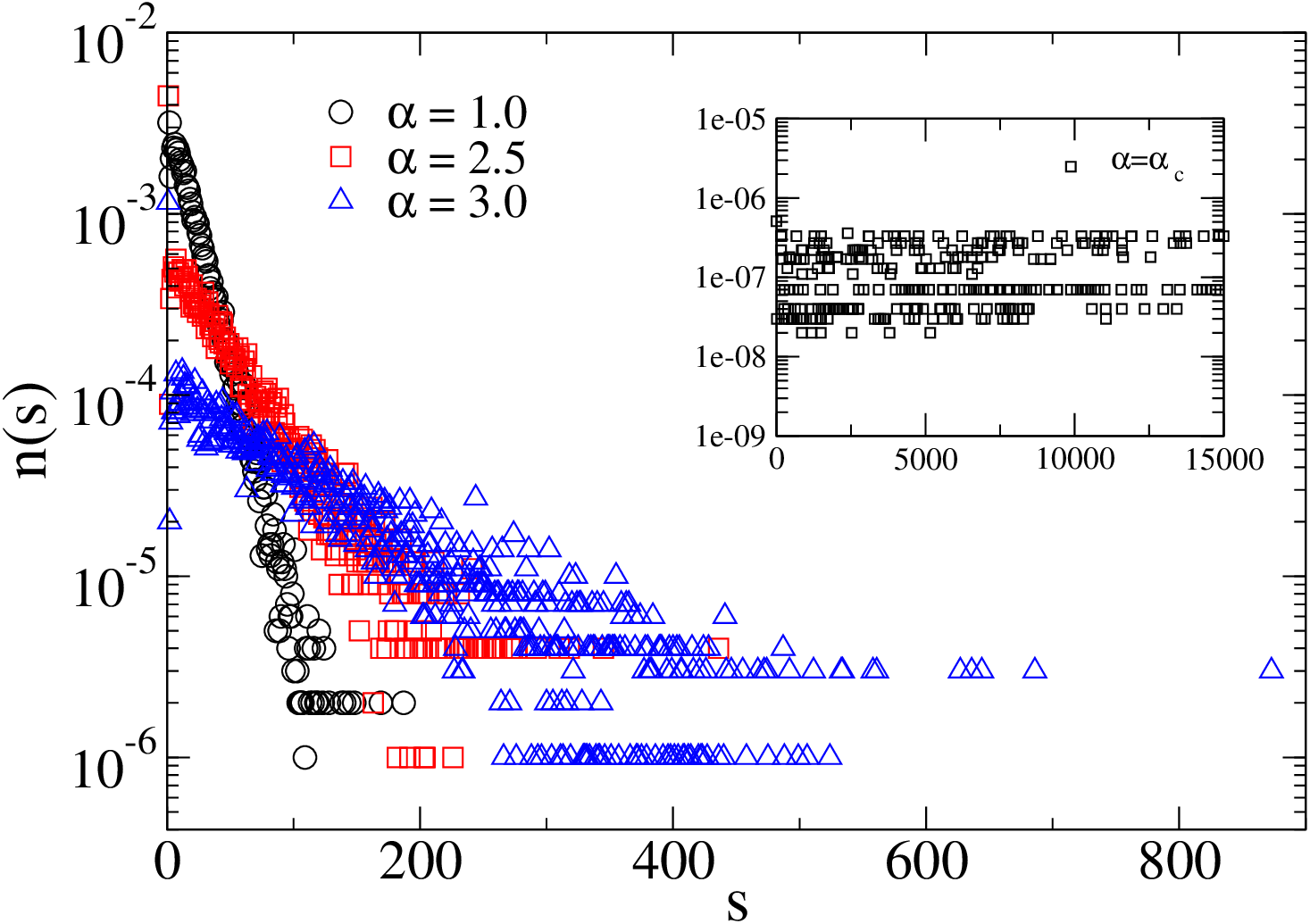}
\end{center}
\caption{The relative cluster size distribution $n(s)$ for some values
  of the parameter $\alpha$ and $N=15000$ nodes. The exponential decay
  of $n(s)$ for every value of $\alpha$ suggests that $\tau=0$, as in
  $1-$dimensional percolation.}
\label{Fig6}
\end{figure}

The analogy with percolation on a ring extends further when we look at
the cluster size distribution (Fig. \ref{Fig6}).  In the limit
$\alpha \to 0$, depending on the number $b$ of bonds added per time
step, there is a critical phase without a giant connected component or
a normal phase with a giant cluster (as predicted by Mendes et al. in
\cite{mendes_pre64}). Since $b=1$ in our case, there is no giant
cluster at $\alpha \to 0$, and the network consists of clusters of
many different sizes $s$, distributed in a pure exponential form
\begin{equation}
\label{eq7}
n(s) \sim e^{-s/\xi(\alpha)}
\end{equation}
\noindent
which suggests that $\tau=0$ \footnote{The general scaling function
  for the distribution of cluster sizes writes $n(s) \sim
  s^{-\tau}e^{-s/\xi(\alpha)}$ with a characteristic cluster size
  diverging near $\alpha_c$ as $\xi(\alpha\to \alpha_c) \sim
  |p-p_c|^{-\nu}$. In $1d$ one has $\nu=1$ and $\tau=0$.}.  Moreover,
as seen in the inset of Fig. \ref{Fig6}, the whole $\alpha>\alpha_c$
phase is critical, in the sense that $\xi(\alpha \ge \alpha_c) =
\infty$.  One might guess this result by noting that when $\alpha\gg
\alpha_c$, the most typical configuration is a line of nodes, that is,
each added node gets connected to its immediate neighbor in the
past. Given $P(a,b)=C (k_a+1)(k_b+1)e^{-\alpha (t_a+t_b)}$, the
probability of joining nodes $a$ and $b$ with ages $t_a$ and $t_b$ and
degrees $k_a$ and $k_b$, respectively, and neglecting, as a first
approximation, terms with contributions smaller than $e^{-4\alpha}$,
the probability that a newly added node gets connected to this line is
$P_{con}\approx
4/27(2e^{-\alpha}+3e^{-2\alpha}+3e^{-3\alpha}+3e^{-4\alpha})$ and a
new cluster emerges with probability $P_{n} \approx (24/27)
e^{-4\alpha}$.  These probabilities get comparable when $4<\alpha<5$
but, since there is always a nonzero chance of not connecting the new
node to the previous ones one finds that for large values of $\alpha$
the system should have scaling properties of a critical
one-dimensional percolation network.

\begin{figure}[!ht]
\begin{center}
\vspace{1.0cm}
\includegraphics[width=0.5\columnwidth,angle=0]{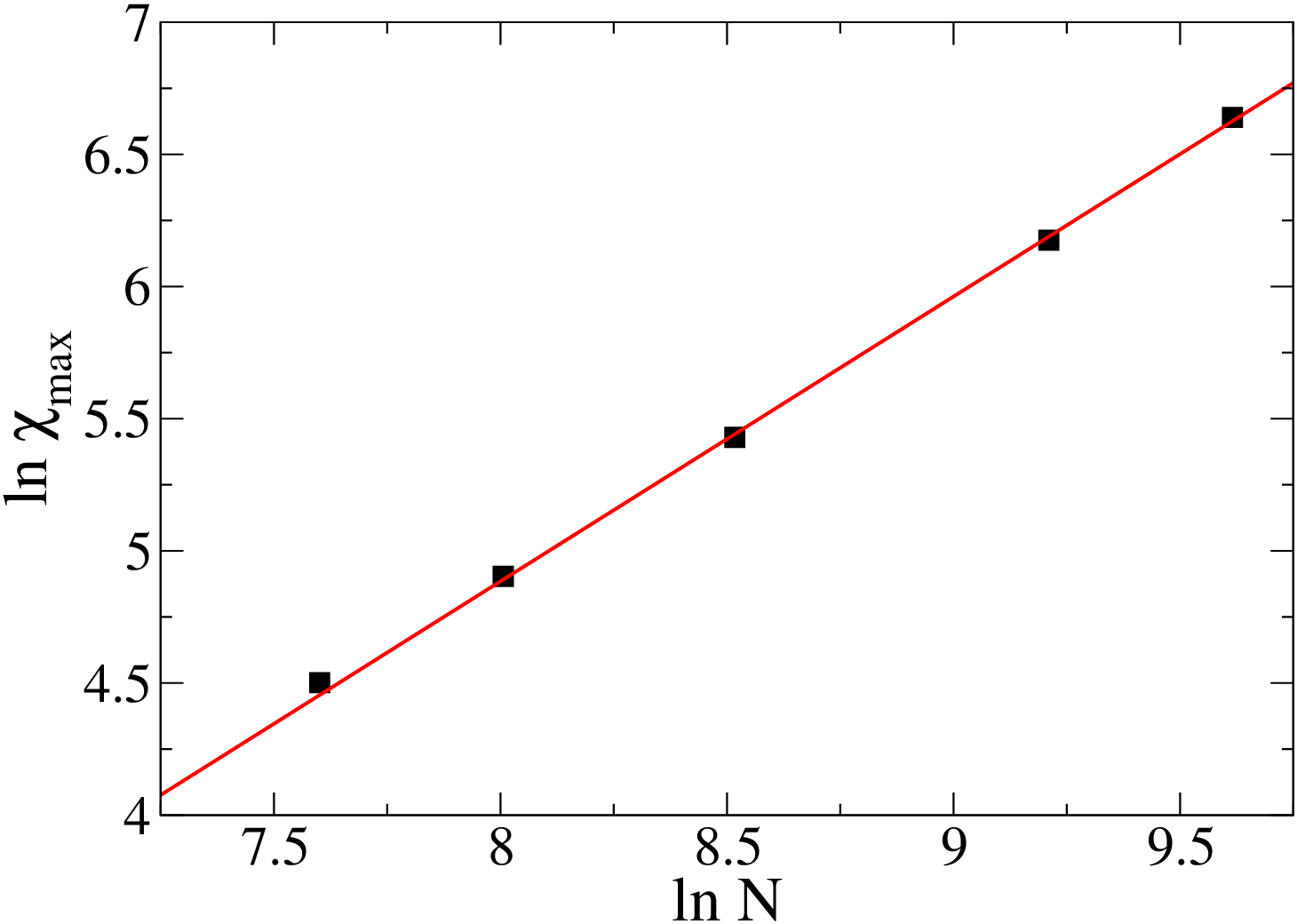}
\end{center} 
\caption{Maximum of the susceptibility $\chi(\alpha_c(N),N)$ versus
  $N$. The best fit indicates a power-law divergence $\chi(\alpha_c)
  \sim N^{\gamma}$ with $\gamma = 1.0(2)$.}
\label{Fig4}
\end{figure}

We also find power-law scaling for the divergence of the
susceptibility at the critical point
\begin{equation}\label{eq4}
\chi(\alpha_c)\sim N^{\gamma},
\end{equation}
\noindent with $\gamma=1.0(2)$, as depicted in Fig. \ref{Fig4}.
Sufficiently close to $\alpha_c$, one can approximate the equation for
the phase transition shift, Eq. (\ref{eq2}), by $\ln
(\alpha_c(\infty)/\alpha_c(N)) \approx
(\alpha_c(N)-\alpha_c(\infty))/\alpha_c(\infty) \sim N^{-1}$, and we find
that $\nu=1$.  This is what one expects at first-order phase
transitions in one-dimensional systems, based on scaling and
renormalization group arguments \cite{fisher}.

\begin{figure}[t]
\begin{center}
\vspace{1.0cm}
\includegraphics[width=0.6\columnwidth,angle=0]{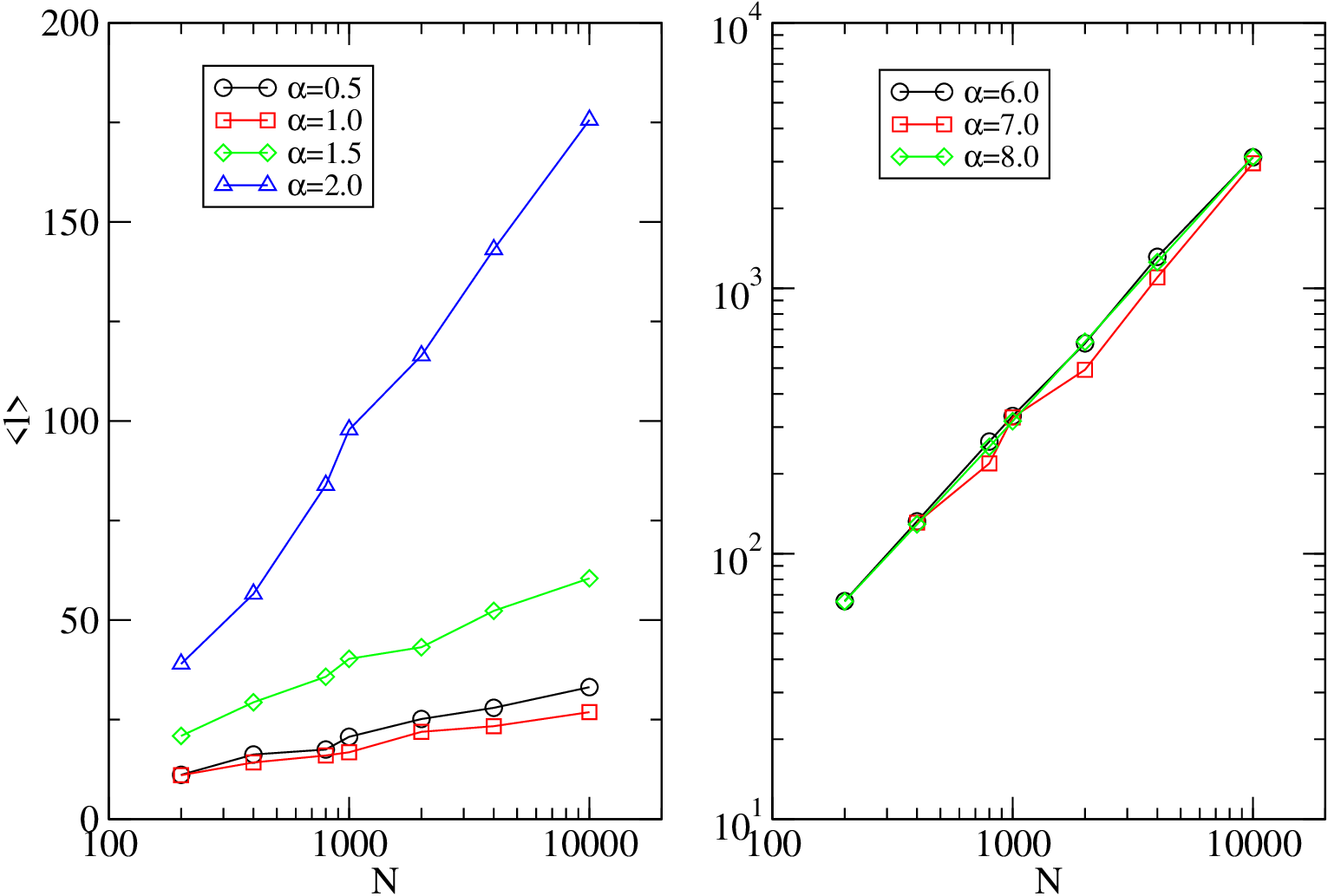}
\end{center}
\caption{Scaling of the average shortest-path between every pair of
  nodes in the largest connected component with system size $N$, for
  values of $\alpha$ below (left panel) and above the critical point
  $\alpha_c$ (right panel). The emergence of a giant connected
  component is followed by a change from tree-like to quasi-one
  dimensional topology, as indicates the crossover from logarithmic to
  linear scaling of $\langle l\rangle$ with $N$.}
\label{fig:lmed}
\end{figure}

In the limit $\alpha \to 0$, when the age-dependence becomes
vanishingly small, links between nodes are added in a random fashion
(although with different rates on nodes with different degrees) and
the finite clusters are locally trees in the large size limit
\cite{mendes_pre64,strogatz}. In this limit one should expect a
logarithmic dependence of the average shortest-path length
$\langle l(N,\alpha)\rangle$ with system size $N$

\begin{equation}
\bar{l}=\frac{1}{N}\sum_{j=1}^{N}\langle l_{ij}\rangle \sim {\cal O}\ln (N),
\end{equation}
\noindent
where $l_{ij}$ is the minimum number of links that must be traversed
to join nodes $i$ and $j$, and $\langle \; \rangle$ means average over
realizations. On the other side, on $d$-dimensional networks $\bar{l}$
is proportional to the linear dimension $L$. Thus, as one increases
the effects of aging in the preferential attachment of links, one
expects a transition from ``small'' to ``large-world'' networks \cite{mendes_epl}, or a change from logarithmic to linear scaling of the average shortest-path
with system size. In Fig. \ref{fig:lmed} we show the average
shortest-path $\langle l(N,\alpha)\rangle$ for every pair of nodes in the
largest cluster of each network generated for different system sizes
$N$. For $\alpha< \alpha_c$ we find logarithmic scaling of $\langle l\rangle$
with $N$, while $\langle l\rangle\sim N$ for $\alpha > \alpha_c$, supporting our
view of a first-order transition in the universality class of $1d$
percolation for this problem.

\section{Conclusions}

We have studied the percolating properties of growing networks with
age and degree preferential attachment of links: nodes introduced
earlier in time are exponentially less likely to acquire new links
than ``younger'' ones and links attach preferentially to nodes with
high degree. One node and one link are introduced at each time step
and the effect of aging on a node with age $a$ and $k$ links is varied
by changing the exponent $\alpha$ in the age- and connectivity-
dependent attachment probability $P(k,a) \propto (1+k)e^{-\alpha a}$.
This model has a discontinuous transition in the relative size of the
largest connected component: below a critical value $\alpha_c$ there
is an extensive number of topologically tree-like clusters of
connected nodes and at a first-order critical point a giant cluster of
linearly connected nodes emerges. We analyzed the properties of this
phase transition with numerical simulations, finding that fluctuations
of the order parameter scale linearly with system size and that, close
to the critical point, the inverse of a characteristic length scale
vanishes linearly with the distance from the critical point,
suggesting that the transition is in the universality class of $1d$
percolation.

\section*{Acknowledgements}

The authors acknowledge financial support from the Brazilian agency CNPq.

\end{document}